\documentclass[twocolumn,aps]{revtex4}

\usepackage{graphicx}
\usepackage{dcolumn}
\usepackage{bm}
\usepackage[section]{placeins}

\begin{document}


\title{Evidence of random magnetic anisotropy in ferrihydrite nanoparticles based on analysis of statistical
distributions}

\author{N. J. O. Silva}
\email{nunojoao@unizar.es}\thanks{Present address: Instituto de
Ciencia de Materiales de Arag\'{o}n, CSIC - Universidad de
Zaragoza, 50009 Zaragoza, Spain}

\author{V. S. Amaral}
\author{L. D. Carlos}

 \affiliation{Departamento de F\'{\i}sica and CICECO, Universidade de Aveiro, 3810-193 Aveiro, Portugal}

\author{B. {Rodr\'{\i}guez-Gonz\'{a}lez}}
\author{L. {M. Liz-Marz\'{a}n}}
 \affiliation{Departmento de Quimica Fisica, Universidade de Vigo, 36310 Vigo,
 Spain}

\author{T. S. Berqu\'{o}}
\author{S. K. Banerjee}
 \affiliation{Institute for Rock Magnetism, University of Minnesota,
 Minneapolis 55455-0128}

 \author{V. {de Zea Bermudez}}
\affiliation{Departmento de Qu\'{\i}mica, Universidade de
Tr\'{a}s-os-Montes e Alto Douro and CQ-VR, Quinta de Prados,
Apartado 1013, 5001-911 Vila Real, Portugal}

\author{A. Mill\'{a}n}
\author{F. Palacio}
\affiliation{Instituto de Ciencia de Materiales de Arag\'{o}n, CSIC
- Universidad de Zaragoza, 50009 Zaragoza, Spain}

\date{\today}

\begin{abstract}
%

We show that the magnetic anisotropy energy of antiferromagnetic
ferrihydrite depends on the square root of the nanoparticles
volume, using a method based on the analysis of statistical
distributions. The size distribution was obtained by transmission
electron microscopy, and the anisotropy energy distributions were
obtained from ac magnetic susceptibility and magnetic relaxation.
The square root dependence corresponds to random local anisotropy,
whose average is given by its variance, and can be understood in
terms of the recently proposed single phase homogeneous structure
of ferrihydrite.

\end{abstract}


\maketitle

\section{\label{sec:Intro}Introduction}

The relation between structural and magnetic properties is of
importance from the point of view of applied and fundamental
research. This relation is not straightforward in systems with
antiferromagnetic (AF) interactions, reduced dimensionality or
size such as nanoparticles. In these systems, surface effects and
disorder play an important role and therefore deviations to the
superparamagnetic (SP) canonic behavior are expected. Such effects
change the relation between anisotropy energy, $E_a$, volume, $V$,
and magnetic moment, $\mu$, found for typical SP systems for which
$E_a$ and $\mu$ are proportional to $V$. This is also the case of
ultrathin films, where anisotropy energy is proportional to
surface area,
leading to perpendicular magnetization. 
Important contribution from surface anisotropy is also found in SP
nanoparticles with ferromagnetic interactions, where surface atoms
constitute a relevant fraction of the total atoms
\cite{FLuis_prb_Co}. Another example of non-proportionality
between $E_a$ and $V$ is two-dimensional Co nanostructures, where
$E_a$ was found to depend on the perimeter
\cite{CoClustersNatMater}. 

The deviations to the proportionality between $V$ and $\mu$ found
in AF nanoparticles are associated to the fact that, in these
systems, the net magnetic moment arises from the uncompensated
and/or canted moments, $\mu_{un}$, that can be present at the
surface, throughout the volume, or both. The relation between
$\mu_{un}$ and $V$ reflect the origin of the moments. In
particular, N\'{e}el has shown that $\mu_{un}$ is proportional to
$V^q$ with $q=1/2$ for moments randomly distributed in the volume,
$1/3$ for moments randomly distributed in the surface and $2/3$
for moments distributed throughout the surface in active planes
\cite{Neel_af1}. In ferritin, a protein where Fe$^{3+}$ is stored
as ferrihydrite, $q$ was estimated to be of the order of $q=1/2$
\cite{Berkowitz_prb, Harris} or to be between $1/2$ and $1/3$
\cite{NJOS_prb} based on magnetization measurements. 
These values were obtained either by using a system with a given
size and estimating the power relation between the total number of
ions and the equivalent uncompensated number of ions
\cite{Berkowitz_prb, NJOS_prb}, or by the usual comparison of
systems with different average sizes \cite{Harris}.
The latter approach is limited by the possibility of synthesizing
identical systems with different average volumes, that usually
covers less that one order of magnitude. An alternative approach
 that takes advantage from the size distribution is developed here.
A sample with a wide distribution can be regarded as one system
containing a set of different average sizes. An approach
reminiscent of this approach was first used by Luis and co-workers
for the determination of the origin of magnetic anisotropy in
gaussian size-distributed Co nanoparticles \cite{FLuis_prb_Co}.
They concluded that surface anisotropy has an important
contribution, since the $E_a$ distribution is narrower than the
$V$
 distribution. The effect of size distributions on the magnetic properties was later
 used to study two-dimensional Co structures by Rusponi et al.
\cite{CoClustersNatMater}. The idea was based on the fact that the
shape of the in-phase component of the ac susceptibility $\chi'$
was critically dependent on the chosen distribution, namely
surface, perimeter and perimeter plus surface distributions. The
authors concluded that perimeter atoms were those relevant to the
reversal process in the Co structures, i. e., $E_a$ depends on the
perimeter. Gilles and co-workers have also tried to use
susceptibility curves to obtain the relation between $\mu_{un}$
and $V$ in ferritin \cite{Gilles_EPJB} but found that their
experimental curves were not very sensitive to the particular
shape of distribution nor the value of $q$ \cite{Gilles_EPJB}.
%
In a different context, the luminosity and the size distributions
of rare-earth-doped nanoparticles were used to establish the
relation between luminosity and size through the size dependent
optical detection probability \cite{casanova_apl}.

In this report we show that lognormal distributed nanoparticle
samples are particularly useful to study the relation between a
physical property and size. This is based in the fact that when
two physical quantities are related by a power function, the power
factor can be readily obtained by comparing the respective
lognormal deviations, due to reproductive properties of the
lognormal distribution function \cite{lognormal_book}. Moreover we
generalize this concept to any distribution function. This method
is of general use and can be simply applied in cases where the
size of the system determined a given physical property by a power
law relation, as in the optical properties of quantum dots
\cite{Brus}. In the present context of magnetism, we apply this
approach to AF ferrihydrite nanoparticles grown in a hybrid matrix
to investigate the relation between $E_a$ and $V$.

%

%

\section{Model}
\label{sec:Model}

\subsection{Relation between distributed quantities}


As pointed out in the previous section, in AF nanoparticles there
is no a priori established relation between $V$ and $\mu_{un}$. At
the same time, $E_a$ and $V$ can also be not proportional. One may
however expect that, in general
\begin{equation}
\label{E_Volume}
 E_a=\alpha V ^p
\end{equation}
where $p$ can be different than 1. In a given situation where the
average values $\langle V \rangle$ and $\langle E_a \rangle$ of
one sample are known it is impossible to determine $\alpha$ and
$p$ simultaneously. Their determination is usually carried out
comparing samples with different $\langle V \rangle$, considering
that $\alpha$ and $p$ are constant in all samples. Here we show
how to determine $\alpha$ and $p$ using magnetic studies on one
lognormal distributed sample.
The probability distribution of $E_a$, $g(E_a)$, is a function of
the $V$ probability distribution $g(V)$:
\begin{equation}
\label{relacao_derivadas} g(E_a)=f(V)/(dE_a/dV)
\end{equation}
If $f(V)$ is a lognormal distribution function with parameters $s_V$
and $n_V$ defined as:
\begin{equation}
\label{lognormal} f(V)=\frac{1}{ V s_V \sqrt{2\pi}}
\exp-\left[\frac{(\log(V/n_V))^2}{{2s_V^2}}\right]
\end{equation}
then $g(E_a)$ is given by:
\begin{eqnarray}
\label{gx_fD} \lefteqn{g(E_a)= \frac{1}{\alpha p (E_a/\alpha)^{(p-1)/p}}\frac{1}{(E_a/\alpha)^{(1/p)} s_V \sqrt{2\pi}}} \nonumber\\
& &
\exp-\left[\frac{[\log((E_a/\alpha)^{(1/p)}/n_V)]^2}{{2s_V^2}}\right]=
\nonumber\\
& &=\frac{1}{ E_a s_E \sqrt{2\pi}}
\exp-\left[\frac{(\log(E_a/n_E))^2}{{2s_E^2}}\right]
\end{eqnarray}
with:
\begin{eqnarray}
\label{relacao_entre_sigmas} n_E=\alpha\, n_{V}^{p} \nonumber\\
s_E=p\,s_V
\end{eqnarray}
%
%
This means that if $V$ presents a lognormal distribution, all
other physical quantities that can be related to $V$ by a power
relation (Eq. \ref{E_Volume}) are also lognormal distributed. More
important, when comparing $V$ and $E_a$, the ratio between the
distribution parameters $s$ is a direct measure of the power $p$,
while the relationship between $n$ values gives information about
$\alpha$. Therefore, the relation between $V$ and $E_a$ in one
sample can be quantitatively derived knowing the lognormal
distribution of $V$ and $E_a$. As one might expect, this method
can be used to determine the relation between any two physical
quantities related by a precise power law similar to Eq.
\ref{E_Volume}.

The relations expressed in Eqs. \ref{gx_fD} and
\ref{relacao_entre_sigmas} are a particular case of the
reproductive properties of the lognormal distribution function
\cite{lognormal_book}. In general, if $X_i$ are independent random
variables having lognormal distribution functions with parameters
$n_i$ and $s_i$ (as defined in Eq. \ref{lognormal}), their product
$Y=c\prod X_{i}^{b_i}$ (with $b_i$ and $c>0$ being constants) is
also lognormal distributed, with $s_Y=\sum b_i s_i$ and $n_Y=c\sum
n_{i}^{b_i}$ \cite{lognormal_book}.
In general, reproductive properties can be used in the analysis of
an output whose inputs are lognormal distributed, as for instance
in quantitative analysis of human information processing during
psychophysical tasks \cite{ieee_lognormal}. However, to the best
of our knowledge, here is the first time that they are used in the
context of physical properties of nanoparticles.


Although many physical properties of interest as size are often
lognormal distributed many others are better characterized by other
functions. This is the case of the anisotropy energy, which is often
described by a gamma distribution \cite{Shliomis_sus_ac,
Jonsson_jmmm, Svedlindh_Palacios_jmmm, FLuis_prb_ferritin}. The
gamma function can be expressed by:
\begin{equation}
\label{gamma} f(x)=\frac{b^{-a}x^{a-1}}{ \Gamma(a)}
\exp-\left(\frac{x}{b}\right)
\end{equation}
with the average of $x$ given by $ab$ and the variance by
$\sigma=ab^2$. For $a>1$, the gamma distribution is similar to the
lognormal function, so that the use of the latter function in the
case where the gamma distribution is more suitable may be a good
approximation. Therefore the use of Eq. \ref{relacao_entre_sigmas}
may also be a good approximation to find $\alpha$ and $p$. These
values may also be found in the general case of a different or an
unknown distribution, by searching for a scaling plot or
numerically \cite{ScaleDist} but, as seen, the validity of the
lognormal distribution makes this task quite straightforward.


\subsection{Anisotropy energy distribution from ac susceptibility and viscosity measurements}
\label{Sec_modelEnergyDist}

The out-of-phase component of the ac susceptibility $\chi''$ is
usually used to obtain the anisotropy energy barrier distribution
of different nanoparticles systems and is given by
\cite{Jonsson_jmmm, Svedlindh_Palacios_jmmm, FLuis_prb_Co,
FLuis_prb_ferritin, FLuis_jmmm}:
\begin{equation}
\label{sus_ac} \chi''(f,T)\simeq\frac{\pi M_{S}^{2}}{6K}k_B T
\ln(1/(f\tau_0))f(E_a)
\end{equation}
where $\tau_0$ is a microscopic characteristic time and $E_a=k_B T
\ln (1/(f\tau_0))$ is the activation energy of the particles
having $\tau$ equal to the characteristic time of measurement
$1/f$. It follows from Eq. \ref{sus_ac} that  $\chi''$ is a
function of $E_a$ and that therefore curves taken at different
frequencies should scale when plotted against  $E_{c}$. At the
same time, $\chi''/T$ is a measure of the anisotropy energy
 distribution $f(E_{c})$. In Eq. \ref{sus_ac} it is considered that the particles
contributing to $\chi''$ at a given $f$ and $T$ are mainly those
with energy equal to $E_a$ \cite{Lundgren_sus_ac} and that the
parallel susceptibility is well approximated by the equilibrium
susceptibility \cite{Shliomis_sus_ac} (i. e. $1/\ln
(1/(f\tau_0))\ll 1$). It is also considered that dipolar
interactions are negligible.

Measurements of the magnetization as a function of time $t$
(viscosity measurements) at temperatures below the blocking
temperature $T_B$ are a complementary way to investigate the
anisotropy energy barrier distribution of different nanoparticles
systems \cite{Labarta_prb, Labarta_jmmm_visco,
Barbara_Ba_Ferrite_jmmm, FLuis_prb_Co}, including ferritin
\cite{St_Pierre_prb}.  With such measurements it is possible to
determine the magnetic viscosity, $S$, defined as the change in
magnetization with $\ln(t)$ of a system held under a constant
applied magnetic field, $h$, and may be written as:
\begin{equation}
\label{viscosity} S(t,T)\equiv\-\frac{\partial M}{\partial \ln
t}=k_B T M_{eq} f(E_a)
\end{equation}
considering that the function $(t/\tau)\exp(-t/\tau)$ is narrower
than the distribution function $f(E_a)$ \cite{Labarta_prb}.
%
%
In ferromagnetic materials $M_{eq}/h=M_{S}^{2}/3K$, where $M_{S}$
is the saturation magnetization and $K$ the anisotropy constant.
It follows directly from Eq.\ref{viscosity} that $S/T$ is
proportional to the anisotropy energy distribution, $f(E_a)$, in
analogy with $\chi''/T$. In fact,
$S$ and $\chi''$ are probing the same energy barrier at different
time scales.

\section{Experimental details}

Ferrihydrite is a low-crystalline AF iron oxide-hydroxide that
typically forms after rapid hydrolysis of iron at low pH and low
temperatures \cite{Ferrihydrite_review}. The structure of
ferrihydrite with domain sizes ranging from 2 to 6 nm was recently
described as a single phase, based on the packing of clusters,
constituted by one tetrahedrally coordinated Fe atom surrounded by
12 octahedrally coordinated Fe atoms
\cite{estrutura_ferrihy_science}. The cell dimensions and site
occupancies change slightly and systematically with average domain
size, reflecting some disorder and relaxation effects. This
picture extends homogeneously to the surface of the domains.
This model contrasts with previous ones, where multiple structural
phases were considered \cite{Drits1,Janney}, and the existence of
tetrahedrally coordinated Fe atoms was a matter of debate
\cite{Eggleton_Fitzpatrick, Manceau_DritsEXAFS}.

The synthesis of the ferrihydrite nanoparticles in the
organic-inorganic matrix (termed di-ureasil) has been described
elsewhere \cite{NJOS_jap}. The particles are precipitated by
thermal treatment at 80 $^\circ$C, after the incorporation of iron
nitrate in the matrix. The sample studied here has an iron
concentration of 2.1 wt\% and was structurally characterized in
detail in Ref. \cite{NJOS_jap_tem}.
M\"{o}ssbauer spectroscopy was measured at selected temperatures
between 4.2 K and 40 K. A conventional constant-acceleration
spectrometer was used in transmission geometry with a $^{57}$Co/Rh
source, using a $\alpha$-Fe foil at room temperature to calibrate
isomer shifts and velocity scale. Ac and dc magnetic measurements
were performed in a Quantum Design superconducting quantum
interference device magnetometer.

\section{Results and Discussion}

\subsection{Relationship between anisotropy energy and size}

\begin{figure}[htb!]
\begin{center}
\includegraphics[width=0.9\columnwidth]{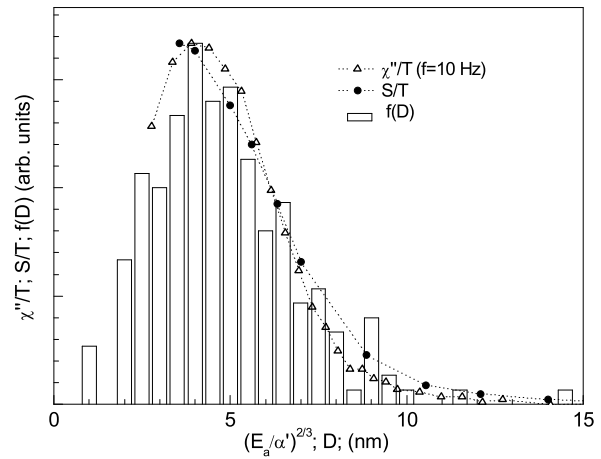}
\end{center}
\caption{\label{dist_raizVolume_fig} Diameter distribution
determined by TEM compared to the anisotropy energy distributions
obtained by $\chi ''/T$ and $S/T$ in a $(E_a/\alpha')^{2/3}$
scale, showing the scaling between $(E_a/\alpha')^{2/3}$ and $D$.
Using other powers $p'$ such as 3, 2, or 1 gives unsatisfactory
scaling.}
\end{figure}

The Fourier transform high resolution transmission electron
microscopy images (FT-HRTEM) and XRD diffraction patterns show the
existence of low crystalline 6-line ferrihydrite nanoparticles.
The nanoparticles are homogeneously distributed, separated from
each other, and have globular habit. The size (diameter, $D$)
distribution can be described by a lognormal function, with $n_D=
4.7\pm 0.2$ nm and deviation $s_D=0.43\pm 0.05$
\cite{NJOS_jap_tem} (see Fig. \ref{dist_raizVolume_fig}). As
expected from the reproductive properties, a lognormal size
distribution results in a lognormal volume distribution.

The in-phase ac susceptibility, $\chi'$, is frequency independent
above $T_F=30$ K. The maxima of $\chi'$ follow a
N\'{e}el-Arrhenius relation:
\begin{equation}
\label{tau} \tau=\tau_0 \exp\left(\frac{E_a}{k_B T}\right)
\end{equation}
The extrapolated $\tau_0$ is of the order of 10$^{-12}$ s, as
found in non-interacting/very weakly interacting nanoparticles
\cite{Dorman_PRB}. As dipolar interactions become relevant, the
extrapolated $\tau_0$ increases. For instance, similar
ferrihydrite/hybrid matrix composites with more concentrated
ferrihydrite nanoparticles (6.5\% of iron in weight), and thus
relevant dipolar interactions, have extrapolated $\tau_0\approx
10^{-17}$ s.

Another evidence of the existence of negligible dipolar
interactions is given by M\"{o}ssbauer spectroscopy results, since
interacting systems have a collapsed V-shaped pattern
\cite{Morup_HypInter_MossAF, Dorman_JMMM_dinamica,
Zhao_Ferrihy_Surface_effectsPRB}. For temperatures around $T_B$
the spectra can be described by the simple sum of a sextet
distribution and a doublet and no signs of a collapsed magnetic
hyperfine field pattern. On the other hand, such collapse is
observed in the ferrihydrite/hybrid matrix sample with 6.5\% of
iron, where dipolar interactions are expected to be relevant.
 At 4.2 K, the M\"{o}ssbauer
spectrum of the sample here studied (2.1\% of iron) shows a
sextet, with a hyperfine field $B_{hf}= 48$ T. This is
characteristic of ferrihydrite nanoparticles low crystallinity, in
accordance with the FT-HRTEM and XRD results.

\begin{figure}[htb!]
\begin{center}\includegraphics[width=0.9\columnwidth]{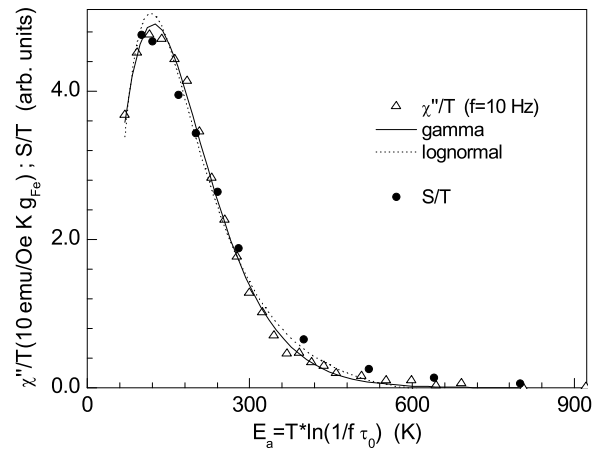}
\end{center}
\caption{\label{distEnergias_fig} Anisotropy energy distributions
obtained by the out-of-phase component of ac susceptibility
($\chi''/T$) and viscosity ($S/T$) measurements. Lognormal and
gamma distribution fits to $\chi''/T$ data are shown.}
\end{figure}

As described in Sec. \ref{Sec_modelEnergyDist}, $\chi ''/T$ and
$S/T$ constitute a direct measure of the anisotropy energy
distribution, observed at different time scales. In Fig.
\ref{distEnergias_fig} we can observe that the distribution
obtained from $\chi ''/T$ and $S/T$ fairly superimpose, meaning
that Eq. \ref{sus_ac} and \ref{viscosity} are good approximations.
Both $\chi ''/T$ and $S/T$ curves are well fitted by a gamma
distribution function, with a=3.3 and b=53 (Fig.
\ref{distEnergias_fig}). As expected for $a>1$, the curves can
also be satisfactorily fitted to a lognormal function, with
$s_{\chi/T}=0.61\pm0.02$ and $\langle \chi/T \rangle= 170\pm4$ K,
and $s_{S/T}=0.65\pm0.02$ and $\langle S/T \rangle= 176\pm4$ K,
respectively. We therefore consider $s_{E}=0.63\pm0.04$ from the
average of $s_{\chi/T}$ and $s_{S/T}$. Since $s_D=0.43\pm 0.05$,
and using Eq. \ref{relacao_entre_sigmas} we directly obtain the
power relation between $E_a$ and $D$, $p'=1.5\pm0.2$, which
corresponds to $p'=3/2$, so that:
\begin{equation}
\label{Rel_Energia_diametro} E_a=\alpha' D^{3/2} \qquad
\textrm{(K)}
\end{equation}

Eq. \ref{relacao_entre_sigmas} can be further used to determine
the proportionality  between $E_a$ and $D^{3/2}$, $\alpha'=18$
Knm$^{-3/2}$. As expected from the above equation, we observe
that, in a $(E_a/\alpha')^{2/3}$ scale, both distributions of
$\chi ''/T$ and $S/T$ superimpose to the diameter distribution
(Fig. \ref{dist_raizVolume_fig}). This is a confirmation that
describing $\chi/T$ and $S/T$ by a lognormal function is a good
approximation for the identification of $p$ and $\alpha$.
Eq. \ref{Rel_Energia_diametro} can be rewritten in terms of the
particle volume as:
\begin{equation}
\label{Rel_Energia_volume}
 E_a=\alpha V^{1/2}
\end{equation}
This means that the anisotropy barriers are randomly distributed
in volume. In each particle, the effective value of $E_a$ is given
by the fluctuation of local $E_a$. This requires that the local
$E_a$ is a random homogeneous quantity. Such homogeneity is
supported by the structure model, since it is composed of a single
phase with in-volume defects, where we can expected $E_a$ to be
locally different.

From Eq. \ref{Rel_Energia_volume} it is still possible to
determine an effective anisotropy energy per volume, $K_{eff}$,
that increases with decreasing $V$, following $V^{-1/2}$. In the 1
-10 nm $D$ range, $K_{eff}$ ranges from $4.7\times 10^{5}$ to $1.5
\times 10^{4}$ J/m$^3$, which are of the order of those found in
the literature \cite{Gilles_EPJB,Pank_reply_comment}. For the
average size of the sample, $K_{eff}=2.9\times 10^{4}$ J/m$^3$.

\subsection{Relationship between magnetic moment and size}

Unlike the case of $E_a$, there is no direct measurement of the
$\mu_{un}$ distribution. A way to obtain this distribution is to
model the dependence of the magnetization with the field $M(H)$ to
a given function of $\mu_{un}$ considering a $\mu_{un}$
distribution. A function usually applied to model $M(H)$ of
nanoparticles is the Langevin function \cite{Berkowitz_prb,
Harris, NJOS_prb, NJOS_jap_tem}. This is a good approximation when
surface effects and anisotropy are negligible. Anisotropy effects
are expected to be relevant in AF nanoparticles due to coupling
between $\mu_{un}$ and the AF axis
\cite{Gilles_EPJB,Morup_jmmm_Aniso}. In AF nanoparticles,
anisotropy effects have been taken into account using a N\'{e}el
(Ising-like) model, considering that $\mu_{un}$ can have only the
AF axis direction \cite{Gilles_EPJB}. On the other hand, recent
simulations show that $M(H)$ is greatly affected by surface
effects, such that a one-spin approach as considered in the
Langevin or N\'{e}el functions are crude approximations
\cite{Kachkachi_efect_anis_MH}.

Despite this situation, we have previously modelled $M(H)$ using a
Langevin distributed function \cite{NJOS_jap_tem} and found that
the parameter $s$ of the $\mu_{un}$ lognormal distribution is
$s_{\mu}=0.9$, so that $q=0.7\pm0.1$ and $\mu_{un}\propto
V^{2/3}$. We note that the value of $q$ here derived is different
to that estimated in Ref. \cite{NJOS_jap_tem} comparing the
average values of the equivalent number of uncompensated ions and
the total number of ions ($q=1/3$). Both $q$ values are obtained
after the same fit procedure performed on the same $M(H)$ curves.
The only difference is the approach for deriving $q$: using
average values of the uncompensated moment and size or using the
information about the distribution of both. This is an example of
how the use of averages may lead to inaccurate estimations, since
the pre-factor of the power law cannot be ignored. In this
scenario the uncompensated moments were to lie on the particles
surface, despite the fact that the energy barriers associated to
the uncompensated moments are randomly distributed in volume.

At this point one should highlight that the Langevin distributed
function may be a too crude description of $M(H)$ to yield a good
estimation of $s_{\mu}$, so that a different scenario is possible:
having no reliable estimation of $s_{\mu}$ we discuss the
situation where the uncompensated moments are associated
(proportional) to the energy barriers, so that a $\mu_{un}\propto
V^{1/2}$, i. e. they are randomly distributed in volume. In fact,
the uncompensated moments are those contributing to the Curie-like
ac susceptibility and those experiencing the blocking phenomena
associated with the onset of $\chi''$ and $S$. Therefore
uncompensated moments should be those relevant in determining the
relation between $E_a$ and $V$. Within this framework, $V^{1/2}$
may be regarded as the equivalent volume that contains the
ferromagnetic-like uncompensated moments. Such relation between
$\mu_{un}$ and $V^{1/2}$ was proposed for antiferromagnetic
nanoparticles by N\'{e}el \cite{Neel_af1} and is consistent with
magnetization measurements performed on ferritin
\cite{Berkowitz_prb, Harris, NJOS_prb}.

\section{Conclusions}

In this report we show that distributed samples can be used to
investigate the relationship between the magnetic anisotropy
barrier $E_a$ and the nanoparticles volume $V$ in a consistent
manner. The relation is accessed by comparing the parameters of
the lognormal distribution of both physical quantities. Size
distribution was obtained by a TEM study and $E_a$ was obtained by
two independent measurements: out-of-phase ac susceptibility and
viscosity measurements. We have applied this method to a
ferrihydrite nanoparticles system and found the relation between
$E_a$ and $V$ in an antiferromagnetic material: $E_a \propto
V^{1/2}$. This shows that the magnetic anisotropy barriers are
randomly distributed in the volume, in accordance to recent
structure studies.

\begin{acknowledgments}

The authors would like to acknowledge E. Lythgoe for critical
reading the manuscript. The financial support from FCT,
POCTI/CTM/46780/02, research grant MAT2004-03395-C02-01 from the
Spanish CICYT. Bilateral projects GRICES-CSIC and Acci\'{o}n
Integrada Luso-Espa\~{n}ola E-105/04, and are gratefully
recognized. N. J. O. S. acknowledges a grant from FCT (Grant No.
SFRH/BD/10383/2002), CSIC for a I3P contract, and financial
support from IRM visiting fellowship program. L.M.L.M.
acknowledges support from the EU Network of Excellence SoftComp.
Institute for Rock Magnetism (IRM) is funded by the Earth Science
Division of NSF, the W. M. Keck Foundation and University of
Minnesota. This is IRM publication \# 0616.
\end{acknowledgments}

\bibliography{bib_NJOSilva}

\end{document}